\documentstyle[preprint,aps]{revtex} % PREPRINT format
\begin{document}
\draft
%..............................................................................
%
\title{Real-space approach to calculation of electric
polarization and dielectric constants}

\author{R. W. Nunes and David Vanderbilt}

\address{Department of Physics and Astronomy,
Rutgers University, Piscataway, NJ 08855-0849}

\date{\today}
\maketitle
%..............................................................................

\begin{abstract}
We describe a real-space approach to the calculation of the
properties of an insulating crystal in an applied electric
field, based on the iterative determination of the Wannier
functions (WF's) of the occupied bands.  It has been
recently shown that a knowledge of the occupied WF's allows
the calculation of the spontaneous (zero-field) electronic
polarization. Building on these ideas, we describe a method
for calculating the electronic polarization and dielectric
constants of a material in non-zero field. The method is
demonstrated for a one-dimensional tight-binding Hamiltonian.
\end{abstract}

%..............................................................................

\narrowtext

Modern electronic band-structure and total-energy
methods, such as those based on the local-density
approximation (LDA) to density-functional theory
\cite{ks_dft}, are not well suited for studying
the properties of materials in an applied electric
field. These approaches typically
rely on solving for the eigenfunctions of the
effective one-electron Hamiltonian. Unfortunately,
the electric field acts as a singular perturbation,
so that the eigenstates of the Hamiltonian are no
longer Bloch states, and the electron band structure
is destroyed,
even for arbitrarily small applied fields.
Linear-response methods \cite{cohen} are capable
of computing derivatives of various quantities with
respect to applied field, but cannot be used to study
the electronic structure of a crystal in a non-zero
electric field directly.
Extensions of these methods beyond linear response
\cite{levine} consist of rather involved expressions
which must be carefully handled in order to avoid
divergences in the static limit.
Formal studies of the structure of the
Wannier-Stark ladder \cite{nenciu} are instructive,
but are not of much practical help from a computational
point of view.

Recently, several groups have introduced new real-space
approaches to the solution of the electronic structure
problem. Based either on the locality of the real-space
density matrix \cite{lnv,daw} or the use of a localized,
Wannier-like representation of the occupied subspace
\cite{mauri,ordej}, these methods were motivated
largely by the search for so-called ``order-$N$ methods''
(for which the computational effort scales only linearly
with system size $N$).  However, these methods also hold
promise for application to the electric-field problem.
The methods based on a Wannier-like representation
\cite{mauri,ordej} appear particularly promising in
view of recent work showing that the electric polarization
of a solid can be directly related to the centers of
charge of the Wannier functions (WF) \cite{king}.

In this paper, we show that the real-space method of Mauri,
Galli, and Car (MGC) \cite{mauri} can be developed
naturally into a practical method for calculating the
response of an insulator to an electric field.
Both the spontaneous and induced electric polarization
$\bf P$ are easily calculated, as are the perturbed charge
density and polarization energy, and dielectric constants
can be obtained by finite differences.
The WF's are expanded in a local basis and truncated to
zero beyond a real-space cutoff radius $R_c$, these being
the only approximations involved.  We apply the method to
a one-dimensional (1D) tight-binding (TB) Hamiltonian and
find that both the spontaneous polarization and dielectric
constant converge quickly, with respect to $R_c$, to the
values obtained by standard $k$-space techniques
\cite{cohen,king}. While the method is demonstrated
in a simple tight-binding context, we see no obvious
obstacles to its implementation in a fully self-consistent
{\it ab-initio} LDA calculation.

We first summarize the MGC scheme using a density-matrix
description.  The exact density matrix of an insulating
crystal is
\begin{equation}
\label{dmdef}
\rho_{\rm exact}=\sum_{n \bf k}\mid\psi_{n \bf k}\rangle
\langle\psi_{n \bf k}\mid\;,
\end{equation}
where the $\{\psi_{n \bf k}\}$ are occupied Bloch
eigenstates of the one-electron (e.g., LDA) Hamiltonian.
For insulators, we can define a unitary transformation
from the $\{\psi_{n \bf k}\}$ to a set of localized
(Wannier-like) functions $\{\psi_j\}$ such that the
occupied subspace is invariant under the transformation.
In terms of these localized orbitals the density matrix is
\begin{equation}
\label{dmrs}
\rho_{\rm exact}=\sum_j^M\mid\psi_j\rangle
\langle\psi_j\mid\;,
\end{equation}
where $j$ runs over a set of localized orbitals that span
the $M$-dimensional occupied subspace of Bloch
eigenstates. The density matrix is spatially localized,
i.e., $\rho({\bf r, r^\prime})\rightarrow 0$ as
$|{\bf r} - {\bf r^\prime}|\rightarrow \infty$, the decay
being exponential for an appropriate choice of phases
of the Bloch functions \cite{kohn_loc}.

In the MGC scheme, Eq.\ (\ref{dmrs}) is taken as an ansatz
for a {\it trial} density matrix, $\rho$.
Each $\psi_j$ is free to vary within a
real-space localization region (LR) of radius
$R_c$, and is set to zero outside the LR.
No orthonormalization constraint is imposed.
The {\it physical} density matrix is then defined by
\begin{equation}
\label{dmtil}
{\tilde\rho}=\rho\sum_{k=0}^K\left (I-\rho\right )^k\;,
\end{equation}
where $K$ is odd. The expectation value of any operator $A$
is then given by ${\rm tr}[{\tilde\rho} A]$.
In the limit where $\rho$ is
idempotent we have ${\tilde\rho}=\rho$, as can be seen from
Eq.\ (\ref{dmtil}). In general, the variarional $\rho$ is
not exactly idempotent, due to the approximations involved.
The parameter $K$ controls the
accuracy of the idempotency requirement \cite{mauri}.
Note that the set $\{\psi_j\}$ in Eq.\ (\ref{dmrs})
spans an $M$-dimensional subspace and that
${\tilde\rho}$ spans the same subspace, since the function
$f_K(x)=1-(1-x)^{K+1}$,
corresponding to
Eq.\ (\ref{dmtil}), is such that $f_K (0)$=0. Moreover,
$f_K (x)$ has its absolute maximum at $x$=1.
{}From these properties it follows that:
(i) when $\rho=\rho_{\rm exact}+O(\delta)$ then
${\tilde\rho}=\rho_{\rm exact}+O(\delta^2)$ ($|\delta|<<1$);
(ii) the total energy, ${\rm tr}[{\tilde\rho} H]$, is
variational provided that all eigenvalues of
the Hamiltonian $H$ are negative.

In what follows, we restrict ourselves to $K$=1, so that
${\tilde\rho}=2\rho-\rho^2$. The variational total energy
is given by $E[\{\psi\}]={\rm tr}[\rho(2-\rho)H]$,
where $H$ has been shifted to make all
eigenvalues negative \cite{explan}. This functional is
minimized with respect to the coefficients of the wave
functions $\{\psi_j\}$ in the given basis. In the limit
$R_c\rightarrow\infty$, minimization of this functional
yields a set of orthonormal orbitals that exactly span the
occupied subspace of $H$, and consequently
$E[\{\psi\}]_{\rm min}=E_{\rm GS}$
(the exact ground-state energy). For a finite $R_c$ the
resulting $\{\psi_j\}$ are not exactly orthonormal and
$E[\{\psi\}]_{\rm min}\geq E_{\rm GS}$, as a result of
the variational nature of the functional.

We next propose a generalized formulation of the MGC
scheme which is more natural and efficient in the case of
periodic systems with small unit cells. In its original
formulation, the MGC scheme is well suited to solving
problems involving large supercells, where a minimization of
a $\bf k$-dependent energy functional, $E^{\bf k}[\{\psi\}]$,
is typically performed only at $\bf k$=0. When applied to
a periodic system with a small unit cell, one would have to
minimize $E^{\bf k}[\{\psi\}]$ independently on a mesh of
$\bf k$-points. In either case, the orbitals $\{\psi_j\}$
are not truly localized, but rather are Bloch functions of
the cell or supercell. Instead, we propose to let the
wavefunctions be truly localized in the manner of Wannier
functions (WF's), $\psi_j=w_{l,n}$ ($l$ and $n$ are cell and
occupied-band indices respectively), and work directly with
the original Hamiltonian.  In this formulation, $\bf k$
plays no role. The electronic degrees of freedom are the
coefficients of the WF's $\{w_{0,n}\}$ in one
cell. The periodicity of the system is now taken into account
by introducing the periodic images of these WF's
in neighboring cells,
\begin{equation}
\vert w_{l,n}\rangle=T_l\,\vert w_{0,n}\rangle \;,
\label{w_trans}
\end{equation}
where $T_l$ is the translation operator corresponding to
lattice vector ${\bf R}_{\,l}$.  In terms of the density
matrix, $\rho=\sum_{l,n} \, \vert w_{l,n}\rangle\langle
w_{l,n}\vert$, the periodicity of the electronic state
is expressed as
\begin{equation}
\label{rho_per}
\rho\left ({\bf r,r^\prime}\right )=\rho\left ({\bf r}
+{\bf R}_{\,l},{\bf r^\prime}+{\bf R}_{\,l}\right ) \;.
\end{equation}

The total energy per unit cell can be written explicitly as
\begin{equation}
\label{E0}
E_0\left [\{w\}\right ]=2\sum_{n=1}^M\langle w_{0,n}
\mid H_0\mid w_{0,n}\rangle -\sum_{n,m=1}^M\sum_l
\langle w_{0,n}\mid w_{l,m}\rangle\langle w_{l,m}
\mid H_0\mid w_{0,n}\rangle\;,
\end{equation}
where $M$ is the number of occupied bands. $E_0$ and $H_0$
denote the total energy and the Hamiltonian at zero electric
field. Since we are dealing with localized orbitals, only a
finite number of orbitals will contribute to the second sum
in Eq.\ (\ref{E0}). Minimization of $E_0$ yields $M$
approximate WF's, for a given choice of $R_c$. Note that
Eq.\ (\ref{E0}) reduces to the original MGC scheme if the sum
in the last term is restricted to $l$=0 only.

It is thus seen that, when applied to a periodic system,
the generalized MGC scheme is equivalent
to the direct determination of its WF's.
This brings us to the work of Refs.\cite{king}, where it
is shown that the electronic contribution to the spontaneous
polarization of a periodic system can be written as
${\bf P}_{\rm e}=-\left ({\rm e}/\Omega\right )\sum_n{\bf r}_n$,
where ${\bf r}_n$
is the center of charge of the WF of band $n$ in zero field,
$e$ is the electronic charge,
and $\Omega$ is the unit cell volume. In the present formalism,
this becomes
\begin{equation}
\label{polar}
{\bf P}_{\rm e}=-\frac{e}{\Omega}\left [2\sum_{n=1}^M
\langle w_{0,n}\mid {\bf r}\mid w_{0,n}\rangle
-\sum_{n,m=1}^M\sum_l \langle w_{0,n}\mid w_{l,m}
\rangle\langle w_{l,m}\mid {\bf r}\mid w_{0,n}
\rangle\right ]\;.
\end{equation}
At this point we have already completed the necessary steps
for a real-space computation of the spontaneous polarization
of a crystalline insulator, by combining Eq.\ (\ref{polar})
with our modified MGC scheme.

Now, we extend the relation between polarization and the
centers of charge of the WF's to a periodic system in
an electric field $\bf F$. The Hamiltonian becomes
\begin{equation}
\label{h_op}
H=H_0+e\,\bf F\cdot\bf r\;,
\end{equation}
Replacing $H_0$ by $H$ in Eq.\ (\ref{E0}) leads to
the total-energy functional
\begin{equation}
E\left [\{w^F\}\right ] = E_0\left [\{w^F\}\right ]
-e\,{\bf F\cdot}{\bf P}_{\rm e}\left [\{w^F\}\right ]\;.
\label {E_F}
\end{equation}
We retain Eq.\ (\ref{w_trans}) and hence Eq.\ (\ref{rho_per}),
and minimize this functional subject only to the constraint of
localization of the field-dependent WF's $\{w^F_n\}$ to the LR,
as before.
Because of the locality of the WF's, the additional terms
which enter Eq.\ (\ref{E_F}) through Eq.\ (\ref{polar})
do not add appreciably to the computational
effort.

We emphasize that our solution does not correspond to the true
ground state of the system.
(There is no true ground state in finite field, as the energy can
always be lowered by transferring charge from ``valence-band''
states in one region to lower-energy ``conduction-band'' states
in a distant region.)
We can think of our solution heuristically as the one which
is generated from the zero-field state by adiabatically
turning on $\bf F$, and keeping the periodicity of the
electronic state expressed in Eq.\ (\ref{rho_per}).
This should be very closely related to what is done when the
field-dependent response of the crystal is measured experimentally.
In order to measure ``static'' properties,
the field must be turned on (or allowed
to oscillate) on a time scale that is slow compared to usual
electronic processes, but fast compared to the
characteristic electronic tunneling rate at the maximum field
encountered.  Thus, the experimental object of study is also
really an excited state (more properly, a very narrow resonance)
of the finite-field Hamiltonian.  Our solution has a very
similar interpretation.

One might object that the existence of a functional $E_0[\{w\}]$,
as in Eq.\ (\ref{E0}), relies on the identification of bands.
In the presence of an electric field, where all gaps disappear
\cite{nenciu}, the existence of bands is not obvious.
Nevertheless, Nenciu \cite{nenciu} has shown
that one can define a sequence of periodic Hamiltonians
$\{H_q\}$, constructed from $H$ by projecting out the
non-periodic part of $\bf F\cdot r$, such that the subspace
of occupied states of a given $H_q$ reduces to that of $H_0$
in the limit $\bf F\rightarrow 0$.  It is argued that
the ``bands'' defined by the Hamiltonian $H_q$ provide an
increasingly accurate description of the finite-field electronic
state as the integer index $q$ is increased.  (Eventually, as
$q$ gets too large, the behavior diverges, in the usual manner of
asymptotic perturbation theory. In other words, the radius of
convergence $F_q$ tends to zero as $q\rightarrow\infty$.)
At least at small $\bf F$, our $\{w^F\}$ are presumably very similar
to the Wannier functions that would be constructed from the bands of
$H_q$, and thus should give a good description of the experimental
electronic state of the system.  We return to this point below.

Before turning to our tests and results, we mention one
technicality.  The minimization of the functional of
Eq.\ (\ref{E_F}) can be performed directly at fixed $\bf F$
using steepest-descent or conjugate-gradient techniques.
This appears to work quite well at weak fields, but can become
unstable for strong fields. Alternatively, one can perform the
minimization with a constraint of fixed $\langle {\bf r}\rangle$
(i.e., fixed $\bf P$), treating $\bf F$ as an adjustable
Lagrange multiplier.
Since $\langle H\rangle-{\bf F}\cdot\langle{\bf r}\rangle$
defines a Lagrange transformation from $\langle H\rangle$ to
$\langle H_0\rangle$, it follows that
$\nabla_{\langle{\bf r}\rangle}\,\langle H_0\rangle=-\bf F$.
In this way, the function $\bf F(P)$ can be mapped out, and
inverted numerically to give $\bf P(F)$.  The latter approach
is more appropriate for investigating the strong-field behavior
of the solutions.

We apply our scheme to a 1D tight-binding 3-band Hamiltonian
in which each unit cell consists of three atoms with one orbital
per atom,
\begin{equation}
\label{ham}
H\left (\alpha\right )=\sum_j\left\{\epsilon_j\left
(\alpha\right )\,c_j^\dagger c_j\,+\,
t\,\left [c_j^\dagger c_{j+1}+h.c.\right ]\right\}\;,
\end{equation}
with the site energy given by
$\epsilon_{3m+k}(\alpha)=\Delta\,cos(\alpha-\beta_k)$.  Here
$m$ is the cell index, $k=\{-1,0,1\}$ is the site index, and
$\beta_k=2\pi k/3$.  This is a simple model of a sliding
commensurate charge-density wave which slides by one period as
the parameter
$\alpha$ evolves through $2\pi$.  For our tests we set $e=1$
and $t=\Delta=-1$, and use $x=\sum_j x_j\,c_j^\dagger c_j$
with $x_j=j/3$ for the TB position operator.  We discuss the
results obtained with only the lowest band filled; the discussion
applies equally to the case of two filled bands (the 3-filled-band
case is trivial).

We first consider the TB Hamiltonian with no external field,
and calculate the spontaneous polarization as a function of
$\alpha$, comparing with the results obtained by the
method of Refs.\cite{king} which we take as exact for
the sake of comparison. The results are shown in
Fig.~\ref{fig:PSxA}(a). In the inset we show the exact
polarization as a function of $\alpha$ in the interval
$[0^{\circ},120^{\circ}]$; also plotted are the results for
$|P_{\rm exact}(\alpha)-P(\alpha)|$ in the same interval, for
$R_c=1.5$ (9 sites within LR) and $R_c=2.5$ (15 sites).
The LR was kept centered at the origin
for all values of $\alpha$. Convergence is already very good
for $R_c=1.5$ with a maximum percentage error of $\sim$1.5\%,
dropping to $\sim$0.5\% for $R_c=2.5$.

We now apply an external electric field to the system.
We minimize Eq.\ (\ref{E_F}) for six fixed values of $F$
between 0.01 and 0.06. In this field region the polarization
$P$ is linear with $F$, to a very good approximation.
In Fig.~\ref{fig:PSxA}(b) we show the linear dielectric
susceptibility $\chi$ as a function of $\alpha$ for $R_c=1.5$,
$R_c=2.5$ and $R_c=3.5$ (21 sites). Also shown are the
linear-response results we obtained using the method of
Ref.\cite{cohen}. $\chi$ converges less rapidly
than $P$, but the maximum percentage
error for $\alpha \in [0^{\circ},120^{\circ}]$
is already $\sim$3.0\% for $R_c=3.5$. Convergence is
systematically worse around $\alpha=60^{\circ}$ where the
WF's are least localized, due to the fact that the gap
between the two lowest bands reaches its minimum value of
0.814 at $\alpha=60^{\circ}$.

Next, we minimize Eq.\ (\ref{E_F}) keeping $\langle x\rangle$
fixed. For a given value of $R_c$ we explore the interval
$[-R_c,R_c]$ of possible values of $\langle x\rangle$.
Fig.~\ref{fig:E_0xP} shows $E_0=\langle H_0\rangle$ as a
function of $P=\langle x\rangle$ for $\alpha=0$ and $R_c$ =
1.5, 2.5, and 3.5. For unconstrained WF's we expect $E_0$ to
be periodic with $P$. Our results reproduce that behavior
remarkably well, except near the boundaries of the LR where
the variational solution becomes poor. Note also that for a
given value of $P$, the value of $E_0$ diminishes with
increasing $R_c$, reflecting the increasing quality of the
variational solution as $R_c$ increases.

We point out that it should be straightforward to
obtain the higher-order dielectric constants from the
$E_0({\bf P})$ curve, by performing careful finite difference
calculations. Moreover, the generalization of our approach to
an {\it ab initio} LDA calculation should be easily implemented.
A localized-orbital basis would be ideally suited, although a
plane-wave basis might also be used \cite{mauri,ordej}.
As in the original MGC scheme, self-consistency can be included
in a straightforward manner, since the density
$n({\bf r})={\tilde\rho}({\bf r},{\bf r})$
remains periodic. Thus the Hartree and
exchange-correlations terms can be computed as usual and do not
contribute to the non-periodic part of the self-consistent
potential.

A final word of caution is in order. It should not be imagined
that there is a well-defined curve $E_0(P)$, periodic in $P$,
which can be obtained by taking the limit $R_c\rightarrow\infty$
of our procedure.  On the contrary, as the LR grows very large,
it becomes possible to construct a solution for $w$ having
arbitrary $\langle x\rangle$ (i.e., arbitrary $P$), and energy
arbitrarily close to $E_0(F=0)$.  This can be done by starting
with the zero-field WF of the occupied band and admixing a small
amplitude (of order $l^{-1/2}$) of a zero-field unoccupied-band
WF at a distance $l$; its energy approaches $E_0(F=0)$ as
$l\rightarrow\infty$. Thus, we have the pathological situation
that $E_0(P)$ becomes perfectly flat in the limit
$R_c\rightarrow\infty$. (When working at fixed $F$, this
pathology shows up in the form of a growing number of false
local minima of the functional of Eq.\ (\ref{E_F}) as
$R_c\rightarrow\infty$.)
However, the Taylor coefficients of $E_0(P)$ (expanded about the
minimum) are well-behaved in the limit $R_c\rightarrow\infty$,
even as its radius of convergence is decreasing to zero.
Underlying this behavior is the asymptotic
nature of the expansion, which is also the case for the
Nenciu \cite{nenciu} construction of the ``polarized'' sub-spaces.
In both cases, full convergence as $R_c\rightarrow\infty$ or
$q\rightarrow\infty$ is obtained only
in the limit ${\bf F}\rightarrow 0$.
We do not claim that our proposed method has superior convergence
or gives more physical results than that of Ref.\ \cite{nenciu}
as $\bf F$ gets large.  But both approaches must have the same
small-field behavior, and hence yield the correct perturbation
coefficients (e.g., linear and nonlinear dielectric constants).
The proposed method also has the advantages of being computationally
tractable and convenient to implement.

In conclusion, we propose a method for calculating the
response of an insulator to an applied electric field
based on a Wannier-function-like representation of the
electronic orbitals. In this approach the spontaneous
polarization, the perturbed charge density and the polarization
energy are easily obtained, and  dielectric constants can be
calculated by finite differences.  The method is variational,
and therefore is well suited to solution by iterative techniques
such as conjugate gradients.  The computational effort scales
only linearly with system size and the method becomes exact as
the cutoff radius used to truncate the Wannier functions is
increased. The method is demonstrated in a simple tight-binding
context, but is also well suited to implementation in a fully
self-consistent {\it ab-initio} LDA calculation.

This work was supported by NSF Grant DMR-91-15342.
R. W. Nunes acknowledges the support from
%the
CNPq - Brazil.
Conselho Nacional de Desenvolvimento Cient\'{\i}fico
Tecnol\'ogico, Brazil.

\vfill\newpage

\begin{figure}
\caption{
Polarization $P(\alpha)$ and linear dielectric susceptibility
$\chi(\alpha)$ for the tight-binding model of the text.
(a) $|P(\alpha)-P_{\rm exact}(\alpha)|$
for $R_c=2.5$ (dashed) and $R_c=1.5$ (dotted line).
Inset: $P_{\rm exact}(\alpha)$.
(b) $\chi (\alpha)$ from linear response (solid line)
and from current method with $R_c=3.5$ (long dashed),
$R_c=2.5$ (dashed), and $R_c=1.5$ (dotted line).
\label{fig:PSxA}}
\end{figure}

\begin{figure}
\caption{
$\langle H_0\rangle$ as a function of $\langle x\rangle$,
for $\alpha=0$.
Solid line, $R_c=3.5$;
dashed line, $R_c=2.5$;
dotted line, $R_c=1.5$.
\label{fig:E_0xP}}
\end{figure}

%-----------------------------------------------------------------------------
\end{document}